\documentclass[twocolumn,showkeys,pre]{revtex4}

\usepackage{epsfig}
\usepackage{dcolumn}
\usepackage{amsmath}
\usepackage{latexsym}
\usepackage{longtable}

\begin {document}

\title
{Structural transitions in aqueous suspensions of natural
graphite}
\author
{Vasily Moraru, Nikolai Lebovka  and Dmitrii Shevchenko}
\affiliation {Biocolloid Chemistry Institute named after F. D.
Ovcharenko, NASU, bulv. Vernadskogo, 42, 03142,Kyiv, Ukraine}
\begin{abstract}
The electric conductivity $\sigma $ and plastic viscosity $\eta $
of aqueous suspensions at different volume fraction of graphite
$\varphi $ and concentration of nonionic surfactant (Triton X-305)
are investigated. The correlations between conductivity and
rheological properties are discussed. A model of structural
transitions in aqueous graphite suspensions is discussed. Two
structural transitions, corresponding to changes in electrical and
rheomechanical properties, are identified as percolation and
sol-gel transitions, respectively. An unusual initial decrease of
the bulk electrical conductivity with graphite volume fraction
$\varphi $ increase was observed in surfactant solution. This fact
is explained as a result of isolating coating formation around the
graphite particles.

\end{abstract}
\keywords{Aqueous graphite suspensions, Electric conductivity;
Viscosity; Percolation transition; Sol-Gel transition}

\maketitle

\section{Introduction}
\label{INTRO}

Colloidal suspensions of carbon have many industrial applications
and are used for production of conducting composite materials,
liquid electrophotographic toners, pigments in inks and paints,
rubber reinforcing and fillers, electrodes and conducting films
[1-7]. The highly dispersed carbon materials (natural graphite,
carbon black etc.) display a strong tendency to agglomerate and
form highly disordered ramified aggregates with the fractal
structures [8-10]. Recently, the properties of these systems are
intensively studied experimentally, but here remain many problems
in understanding of correlations between morphology of the
aggregates and physical properties. Especially, we may note the
practically important problem related with electrical percolation
behavior of the carbon-based composite materials [11-15].

In this work we studied the structural transitions in the aqueous
suspensions of natural graphite in the presence of nonionic
surfactant by means of electrical conductivity and viscosity
measurements. Because of high hydrophobicity of graphite, it
particles tend to aggregate in water suspensions and to form a gel
phase, but addition of a surfactant allows to achieve fine
adjustment of aggregation processes through changes in surfactant
concentration.

\section{Materials and methods}

\subsection{Materials}

\begin{figure}
\centerline{\epsfxsize=7.0cm \epsfbox{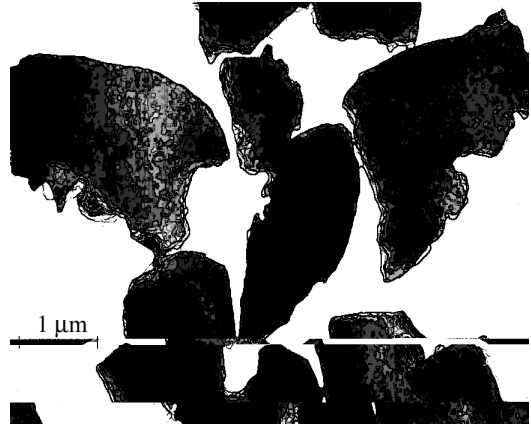}}
\caption{Transmission electron micrograph of the natural graphite
sample.} \label{f01}
\end{figure}

The natural graphite (product of Zavalie Plant, Ukraine, carbon
content 99.5\%, type C-0) was chosen for investigation. An example
of the transmission electron micrograph of the natural graphite
sample is presented in Fig. 1. The electron microscopy data
confirm the presence of particles with plate-like geometry. The
granulometric composition of dispersions was studied by laser
diffraction microanalyzer Analysette-22 (Fritsch). The
distribution curves cover the size range 1-10 \quad $\mu $m \quad
and have a narrow maximum at $ \approx $5 $\mu $m
(Fig.2).

\begin{center}
\textbf{Table.} Some properties of the natural graphite
\end{center}
\begin{longtable}
{|p{190pt}|p{40pt}|} a & a  \kill \hline Property&
Value \\
\hline Diameter of particle, \textit{d} ($\mu $m)&
$ \approx $5 \\
\hline Plate thickness, \textit{h} ($\mu $m)&
0.1-0.5 \\
\hline Particle aspect ratio, \textit{d/h}&
10-50 \\
\hline Specific surface area, \textit{S} (m$^{2}$/g)&
20 \\
\hline Exchange capacity, \textit{q} (meq/g)&
0.05 \\
\hline Surface charge density (pH=10) , $\sigma _{s}$ (C/m$^{2}$)
&
0.12 \\
\hline Specific heat wetting, \textit{w} (J/m$^{2}$)&
0.09 \\
\hline
\end{longtable}

\begin{figure}
\centerline{\epsfxsize=7.0cm \epsfbox{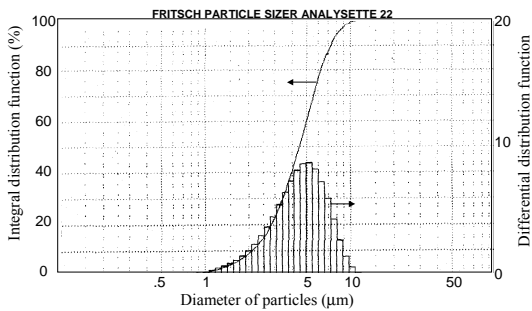}}
 \caption{Integral
and differential distribution functions of particle diameters $d$
for the natural graphite. } \label{f02}
\end{figure}

The Table presents the main properties of our sample of natural
graphite. Here, the value of the specific surface area \textit{S}
was determined from the water vapor adsorption isotherms using the
vacuum microbalance apparatus and from the dye (methylene blue)
adsorption isotherms in water solutions. The exchange capacity,
\textit{q} and the surface charge \textit{$\sigma $}$_{s}$ were
determined using the conductometric and potentiometric titration
of dispersions by 0.1 M KOH solution and 0.2 M KCI solution. The
heat of wetting \textit{w} was determined with the help of DAK-1M
microcalorimeter (Pribor, Moscow).

We used Triton X-305,
CH$_{3}$-C(CH$_{3}$)$_{2}$-CH$_{2}$-C(CH$_{3}$)$_{2}$-C$_{6}$H$_{5}$O-
(CH$_{2}$CH$_{2}$O)$_{n}$H, where n=30, as a non-ionic surfactant.
The background electrolyte concentration of 0.01M KCl was present
in all the cases. The stated values of pH and ionic strength of
dispersions were adjusted by the combination of 0.01 M HCl, KCl
and KOH solutions. The water, used in this study, was double
distilled and had conductivity 5x10$^{-6}$ S/cm at 20°C. All the
reagents were of analytical grade. The suspensions were prepared
by sonicating (UZDN-2T sonifier, Russia) during 1 minute. The
sonifier was operated at a frequency of 22 kHz, with the output
power 150 w. Then, after cooling, the sample was placed into the
thermo-stabilized cells for measurements of conductivity and
rheological parameters.

\subsection{Experimental methods}

The electrical conductivity, zeta potential and rheological
parameters of natural graphite in aqueous suspensions were
measured at fixed temperature T=293 K.

\subsubsection{Electrical conductivity}

The electrical conductivity measurements were made using Ð-5021
(Tochelectropribor, Kiev, Ukraine) at the frequency of
\textit{f}=1 kHz in a cell with platinum electrodes with the
surface area of 1 cm$^{2}$

\subsubsection{Zeta potential}

The zeta potential measurements were made using
microelectrophoresis technique at the electric field strength
\textit{E} = 5-6 V/cm and the volume fraction of graphite
particles $\varphi  \sim $10$^{-4}$. The average of, at least,
three measurements for each sample was recorded. No corrections
were made for polarization and relaxation effects of the
electrical double layer, as far as in our case the condition
\textit{d/$\lambda $} $ \ge $ 100 (where \textit{d} is the average
particle diameter and $\lambda $ is Debye screening length) is
fulfilled [16].

\subsubsection{Rheological behavior}

The viscosity measurements were made using the coaxial cylinder
viscometer of Rheotest-2 model (Germany) for shear rates $\mathop
{\gamma} \limits^{ \bullet}  $ from 0 to 1312 s$^{-1}$.

\subsubsection{Influence of pH}

The pH values of suspensions were measured using the pH meter with
glass electrodes EV-74 (Electrotochpribor, Kiev). It should be
noted that pH of is an important parameter, which controls the
structure of graphite suspension. The graphite surface displayed
weak $\pi $-basic properties owing to the presence of $\pi
$-electrons, delocalized over the entire macroaromatic skeleton.
From the other side, the graphite surface displays acidic
properties owing to the presence of small quantity of the
chemically grafted oxygen-containing groups. As a result, the
natural graphite always displays properties of a typical
ampholyte. The isoelectric point, where the zeta potential is
zero, $\zeta $=0, was found to be at pH$_{iep}\approx
$4.5-5.5 (Fig.3).
So, the graphite colloidal particles have a
positive charge at pH$<$pH$_{iep}$, and a negative charge at pH
$>$pH$_{iep}$. This strong dependence of the acid-basic and
electrical surface properties of graphite from pH values results
also in the structural changes of graphite suspensions with pH. As
an example, we may refer to the fact that the plastic viscosity of
graphite suspension $\eta$ may be an extreme function of pH
(Fig.3). In this work, all the investigations were done at the
fixed value of pH = 6.3.
\begin{figure}
\centerline{\epsfxsize=7.0cm \epsfbox{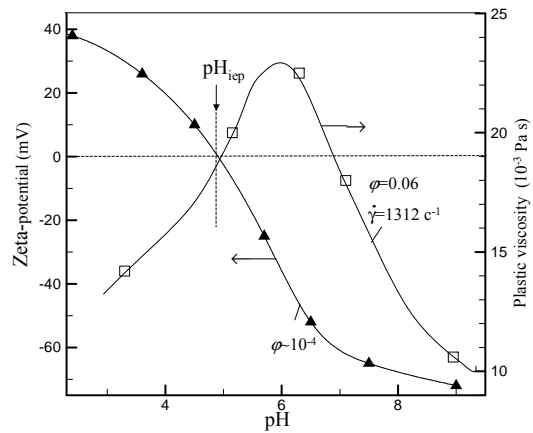}}
\caption{Electrokinetic zeta-potential and plastic viscosity $\eta
$ as a function of pH.} \label{f03}
\end{figure}

\section{Results and discussion}

Electrical conductivity and rheological properties of suspensions
are very sensitive to their structural organization, which is
controlled by the balance of interparticle forces [17]. An
important information about the structural transitions and
mechanisms of particle aggregation in water suspensions may be
obtained from analysis of experimental data on these
characteristics
\begin{figure}
\centerline{\epsfxsize=7.0cm \epsfbox{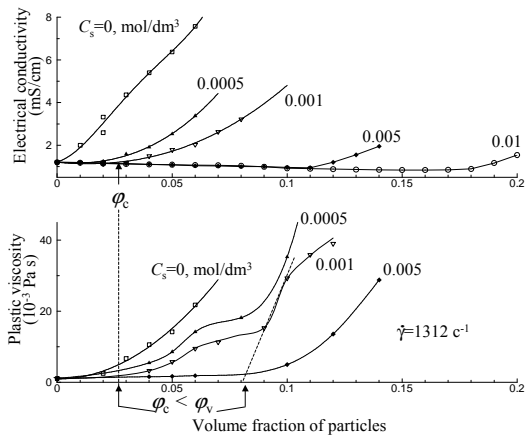}}
 \caption{Plots of
the electrical conductivity $\sigma $ and plastic viscosity $\eta$
of suspensions versus volume fraction of graphite particles
$\varphi$ at different Triton X-305 concentrations $C_{s}$.}
\label{f04}
\end{figure}
Figure 4 shows electrical conductivity $\sigma $\textit{}  and
plastic viscosity $\eta $, measured at shear rate $\mathop
{\gamma} \limits^{ \bullet}=1312$ s$^{-1}$ versus volume fraction
of the natural graphite $\varphi $ in water at different
concentrations C$_{s}$ of Triton X-305. In the absence of any
surfactant, the electrical conductivity $\sigma $\textit{ }and
viscosity $\eta $ increase sharply even at very small volume
fractions $\varphi\approx  0.01-0.02$ (1-2 vol.\%). It
reflects existence of a strong aggregation between the carbon
particles and formation of a highly interconnected network between
the particles of anisotropic geometry [18-20]. In the presence of
a surfactant, the fluidifying effect is observed, It results from
decrease of the aggregation of particles. The surfactant molecules
form the stabilizing layers on the surfaces of the graphite.

\begin{figure}
\centerline{\epsfxsize=7.0cm \epsfbox{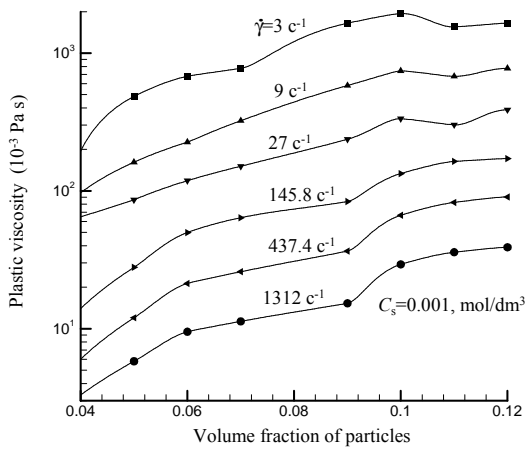}}
 \caption{Plots of
the plastic viscosity of suspensions $\eta $ versus volume
fraction of the natural graphite particles $\varphi $ measured at
different $\mathop {\gamma} \limits^{ \bullet} $.} \label{f05}
\end{figure}

Because of a high electrical contrast of these systems (specific
conductivity of graphite particles is of an order $\sigma
 \approx 10^{3}-10^{5}$ S/cm [21] and is
very high as compared with conductivity of water solutions,
$\sigma \approx 10^{-3}$ S/cm ) the typical percolation behavior
of conductivity $\sigma$ is observed [22]. A considerable increase
of the conductivity of suspensions begins only after the certain
threshold value of $\varphi =\varphi_{c}$ is exceeded (Fig. 4).
The similar behavior was observed for other highly conductive
carbon-based composites [8,10,11,13,15,18,19]. The viscosity of
suspensions continuously decreases with volume fraction $\varphi $
increase. However, as rule, the observed $\eta$ versus $\varphi$
dependencies are step-like (Fig.4). The viscosity begins to
increase sharply only at concentrations $\varphi>\varphi_{v}$, and
$\varphi_{v}$ value is noticeably high than value of
$\varphi_{ñ}$. As surfactant concentration $C_{s}$ increases, both
threshold values, $\varphi_{c}$ and $\varphi_{v}$, also increase.
For example, in Fig. 4, two concentrations of $\varphi_{ñ}$ and
$\varphi_{v}$ are shown for the case when $C_{s}=0.001$
mol/dm$^{3}$. The similar step-like behavior $\eta(\varphi$) was
observed also for other values of $\mathop {\gamma} \limits^{\ast}
$ (Fig. 5).
\begin{figure}
\centerline{\epsfxsize=7.0cm \epsfbox{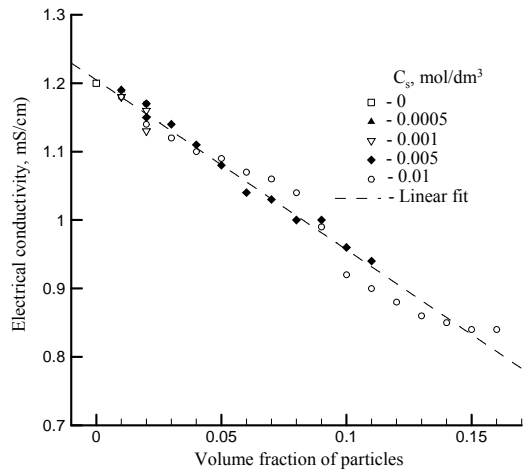}}
 \caption{Initial
linear decrease of the electrical conductivity of suspensions
$\sigma $ with increase of the volume fraction of natural graphite
particles $\varphi $.} \label{f06}
\end{figure}
In the region where $\varphi$ do not exceed the threshold value
$\varphi_{c}$, the initial linear decrease of the bulk electrical
conductivity $\sigma$ with $\varphi$ increase is observed (Fig. 6)
\begin{equation}
\sigma(\varphi)=\sigma_i=\sigma(0)(1 -\alpha\varphi),\text{\quad
 at\quad} \varphi\le\varphi_{c}.\label{Eq1}
\end{equation}
Here, $\sigma(0) = 1.23$ mS/cm is the conductivity of the 0.01 N
aqueous solution of KCl at $T=293$K and $\alpha =2.05$ is the
fitting parameter.

This behavior is rather unusual, as far as electrical conductivity
of the system decreases with increasing of the ratio of more
conductive component. In fact, Maxwell obtained first the Eq. (1)
for dilute dispersions of isolating spherical inclusions in a
conductive host matrix [23]. It is valid for $\varphi<0.1$ and the
coefficient $\alpha$ is equal to $1.5.$

The initial decrease of electrical conductivity may be interpreted
in view of the low effective conductivity of carbon particles in
the presence of surfactant as compared with the water solution
matrix. The larger slope $\alpha=2.05$ as compared with Maxwell
approximation may be explained by existence of the particle size
distribution and anisotropy of the particle geometry. When
concentration of the particles exceeds a certain threshold value
($\varphi\ge\varphi_{c}$), which depends on the surfactant
content, a usual scaling percolation law is observed
\begin{equation}
\sigma= \sigma _{i}+\beta (\varphi-\varphi_{c})^{t}, \text{\quad
at \quad} \varphi> \varphi _{c},\label{Eq2}
\end{equation}
\noindent where $\beta$ is a parameter and $t$ is the critical
index of conductivity.

The value of the critical conductivity index $t$ was estimated
from the best fits to data and it was found to be close to $t
\approx 2.0$, which corresponds to a classical random percolation
case [22]. It means that long-range correlations, able to change
universality of a short range random percolation, are absent at
the percolation threshold. It should be noted that measured values
of $t$ varied within a wide interval for composites of
carbon-black [15].

\begin{figure}
\centerline{\epsfxsize=7.0cm \epsfbox{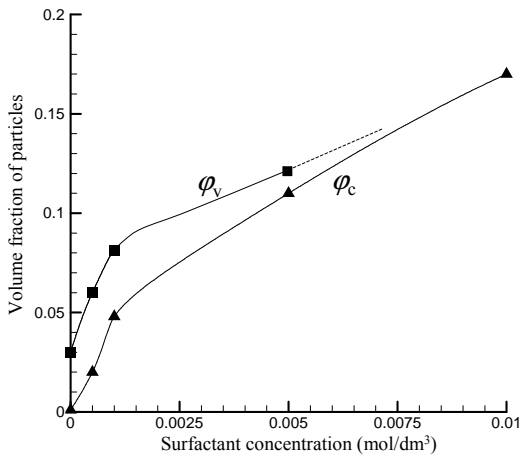}}
 \caption{Critical
volume fractions of particles $\varphi _{c}$ and $\varphi _{v}$
(estimated from measurements of electrical conductivity and
viscosity, respectively) versus surfactant concentration $C_{s}$.}
\label{f07}
\end{figure}

The concentrations of the first and second structural transitions
$\varphi_{c}$ and $\varphi_{v}$ versus surfactant concentration
\textit{C}$_{s}$. are presented in Fig. 7. As surfactant
concentration $C_{s}$ increases, the threshold values of
$\varphi_{c}$ and $\varphi_{v}$ also increase. At high enough
surfactant concentration, the two-step form of the $\eta$ versus
$\varphi$ curve disappears, and here $\varphi_{ñ}\approx
\varphi_{v}$.

The complex behavior observed for electrical conductivity $\sigma
$ and plastic viscosity $\eta $ evidences existence of several
structure levels in suspensions and reflects specific features of
coagulation and thixotropic properties of these suspensions. The
three-dimensional network structures of disperse systems may
display different topological properties, depending on the volume
fraction of the solid phase, geometry of particles and presence of
the chemical additives and structure-forming agents [24]. In
concentrated suspensions, different structures may arise, such as
periodical colloid structures and random statistical networks
formed from chains of aggregated particles, or compact aggregates
linked by coagulation chains.

The structural transitions at $\varphi_{c}$ and $\varphi_v$
correspond to changes in electrical (conductivity $\sigma$) and
rheomechanical (viscosity $\eta$) properties of a system, and we
believe that they may be identified as percolation and sol-gel
transitions, respectively.

An illustration of our model of structural transitions in aqueous
suspensions of graphite is presented in Fig. 8. At low
concentration of solid phase, the long-range type mechanism of
structure formation prevails. We can assume that surfactant
molecules form nonconductive films on the hydrophobic surface and
stabilize a graphite suspension through modification of the
hydrophilicity and zeta potentials of colloidal particles. These
films have low conductivity, and the coated graphite particles
behave effectively as nonconductive. With $\varphi $ increase at
given $C_s$ the isolation of particles destroys. At the critical
concentration $\varphi $ = $\varphi_{c}$ an infinite percolation
cluster of the conducting graphite particles appears for the first
time. The critical concentration $\varphi_{v}$ corresponds to the
critical concentration of the structure formation. At this point
$\varphi $ = $\varphi_{v}$, an infinite cluster with non-local
elastic properties appears for the first time, and this point may
identified as the sol-gel transition point.

\begin{figure}
\centerline{\epsfxsize=7.0cm \epsfbox{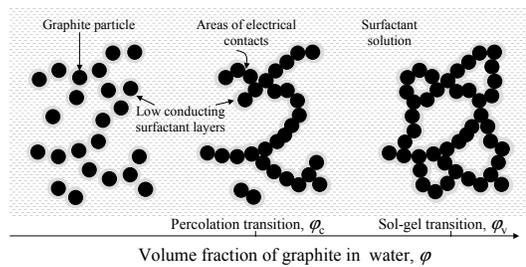}}
\caption{Illustration of a model of structural transitions in
aqueous graphite suspensions. Black circles represent the carbon
particles and gray shells represent the low-conducting surfactant
layers.} \label{f08}
\end{figure}

\section{Conclusions}
\label{CONCLUSIONS}

 Experimental data on electric conductivity and
plastic viscosity of aqueous natural graphite suspensions as
function of solid volume fraction and nonionic surfactant
concentration were presented. Electric conductivity and plastic
viscosity displays different sensitivity to structural transitions
in suspensions. Two different concentration points, corresponding
to changes in electrical and rheomechanical properties, were
identified as percolation and sol-gel transitions, respectively.
We obtained the critical index value as close to t$ \approx $2.0,
which corresponds a classical random percolation case. The
surfactant molecules may form isolating coating around particles
of conductive graphite, and it results in initial decrease of
suspension conductivity with increase of solid content.

\section*{Acknowledgements}
This material is based upon work partially supported by the
National Academy of Science of Ukraine under the program
``Nanosystems, Nanomaterials and Nanotechnologies'' and Grants No.
2.16.1.4 (0102V007058) and 2.16.2.1(0102V007048). Authors also
thank Dr. N.S. Pivovarova for her help with preparation of the
manuscript.

\end{document}